\documentclass[english,pra,preprint,superscriptaddress]{revtex4-1}
\usepackage{amsmath,amssymb,bm}
\usepackage{graphicx}
\usepackage{dcolumn}
\usepackage{ifthen}
\usepackage{babel}
\usepackage{color}
\usepackage{hyperref}
\hypersetup{linktoc=page}
\hypersetup{
    colorlinks,
    citecolor=blue,    filecolor=blue,
    linkcolor=blue,     urlcolor=blue
}

\begin{document}
\title{Quantum erasing of laser emission in N$_2^+$}
\author{Rostyslav Danylo}
\affiliation{Laboratoire d'Optique Appliqu\'ee, ENSTA ParisTech, CNRS, Ecole Polytechnique, Institut Polytechnique de Paris, 91762 Palaiseau, France}
\affiliation{Shanghai Key Lab of Modern Optical System, University of Shanghai for Science and Technology, 200093 Shanghai, China}
\author{Guillaume Lambert}
\affiliation{Laboratoire d'Optique Appliqu\'ee, ENSTA ParisTech, CNRS, Ecole Polytechnique, Institut Polytechnique de Paris, 91762 Palaiseau, France}
\author{Yi Liu}
\affiliation{Shanghai Key Lab of Modern Optical System, University of Shanghai for Science and Technology, 200093 Shanghai, China}
\author{Vladimir T. Tikhonchuk}
\affiliation{University of Bordeaux-CNRS-CEA, CELIA, UMR 5107, 33405 Talence, France}
\affiliation{ELI-Beamlines, Institute of Physics, Czech Academy of Sciences,  25241 Doln\'{i} B\v{r}e\v{z}any, Czech Republic}
\author{Aur\'{e}lien Houard}
\affiliation{Laboratoire d'Optique Appliqu\'ee, ENSTA ParisTech, CNRS, Ecole Polytechnique, Institut Polytechnique de Paris, 91762 Palaiseau, France}
\author{Andr\'e Mysyrowicz}
\email{andre.mysyrowicz@ensta.fr}
\affiliation{Laboratoire d'Optique Appliqu\'ee, ENSTA ParisTech, CNRS, Ecole Polytechnique, Institut Polytechnique de Paris, 91762 Palaiseau, France}

\date{\today}

\begin{abstract}
Cavity-free lasing of N$_2^+$ induced by a femtosecond laser pulse at 800~nm is nearly totally suppressed by a delayed twin pump pulse. We explain this surprising effect within the V-scheme of lasing without population inversion. A fast transfer of population between nitrogen ionic states X$^2\Sigma_g^+$ and A$^2\Pi_u$, induced by the second pulse, terminates the conditions for amplification in the system. Appearance of short lasing bursts at delays corresponding to revivals of rotational wave packets are explained along the same lines. 
\end{abstract}

\maketitle

\section{introduction}

According to quantum mechanics, electronic transitions are described not by probabilities but by \textit{amplitudes} of probabilities. This leads to the possibility, when two electronic levels are coupled to a common third level by coherent radiation fields, of interference effects with unusual consequences. Counterintuitive effects such as suppression of spontaneous emission \cite{Harris1997}, induced transparency of an opaque medium \cite{Alzetta1976, Arimondo1996} or lasing without population inversion \cite{Harris1989, Svidzinsky2013} have been reported. Recently, we have interpreted the cavity-free lasing of N$_2^+$ pumped by femtosecond laser pulses as being due to such a lasing without population inversion \cite{Mysyrowicz2019, Tikhonchuk2020}. Here we report a peculiar consequence from quantum interference in this system, namely the quenching of this laser emission when a second delayed femtosecond pump pulse is injected in the gas.

\section{results and discussion}
\label{sec1}

\begin{figure}[!ht]
\centering
\includegraphics[width=10cm]{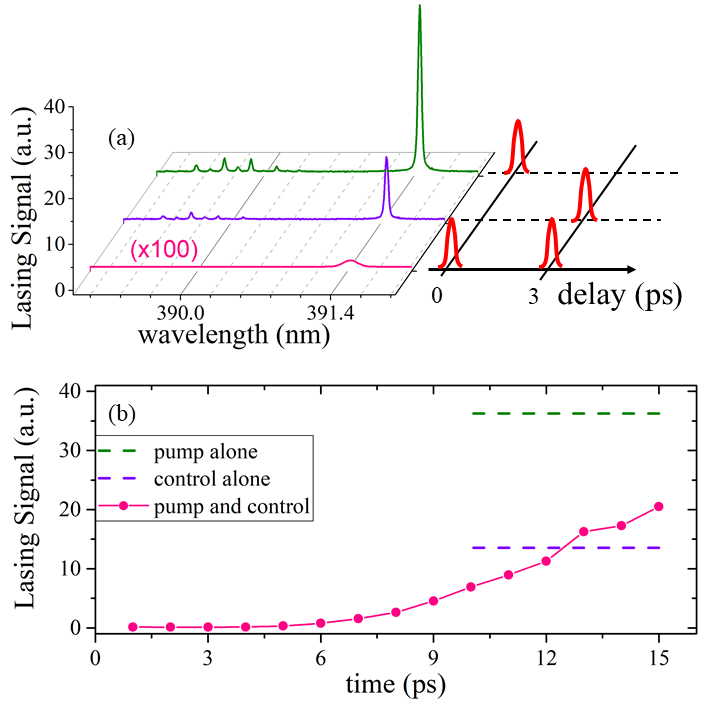}
\caption{a) Time-integrated lasing signal in the forward direction obtained with a 800~nm pump pulse of 45~fs duration, NA=0.0125. Upper trace (green curve): pump pulse only (1~mJ energy), middle trace (violet curve): delayed control pulse only (1~mJ energy), lower trace (pink curve): consecutive pump and control 1~mJ pulses with 3~ps delay. b) Time--integrated lasing signal at 391.4~nm plotted as a function of delay \textit{dt} between the pump and control pulses. The N$_2$ gas pressure is 30~mbar.}
\label{figure1}
\end{figure}

The erasing effect is shown in Figure \ref{figure1}. A laser pump pulse of 45~fs duration and 1 mJ energy with central wavelength at 800~nm is focused with a 40 cm lens (NA$~=0.0125$) in a 1~m length gas chamber filled with nitrogen gas at low pressure (30~mbar). It leads to the formation of a plasma channel about 1~cm in length and to cavity-free self-seeded lasing at 391.4~nm in the direction of the pump pulse (Figure \ref{figure1}a, upper trace). The 391.4~nm wavelength corresponds to a transition between second excited state B($\nu=0$) and ground state X($\nu=0$) of N$_2^+$ \cite{Liu2013,Yao2013,Yao2016,Xu2015,Li2019,Ando2019,Richter2020,Arissian2018,Britton2019,Liu2015,Britton2018}. As expected, another pulse, hereafter called control pulse, of same energy and duration gives a signal of same order (Figure \ref{figure1}a, middle trace). However, when both pump and control pulses are sent consecutively within a few picoseconds, one observes a reduction by a factor $\sim1000$ of the global lasing output, as shown in Figure \ref{figure1}a, lower trace. This surprising effect is robust, as shown in Figure 1b, where the time-integrated lasing signal at 391.4~nm is plotted as a function of time between the two 800~nm pulses. As can be seen, the erasing effect lasts for more than 5~ps, much longer than the pump or control pulse duration.\par 
Before discussing this result, it is useful to describe the properties of 800~nm femtosecond pulse-induced N$_2^+$ lasing and its interpretation. Cavity-free lasing in the forward direction happens within the plasma column created in air or pure nitrogen gas by an intense femtosecond laser pulse with peak intensity $\sim$10$^{14}$~W/cm$^2$. The emitted lasing pulse at 391.4~nm is retarded from the pump pulse by several picoseconds and its duration varies between 1-2~ps and 10~ps or more near lasing threshold, depending on pressure \cite{Liu2015}. This behavior is typical for the collective emission of many emitters locked to a common phase \cite{Gross1982}. Injection of a weak femtosecond seed pulse at 391~nm amplifies the signal by 2-3 orders of magnitude \cite{Yao2013,Yao2016,Arissian2018,Liu2015}. Three models have been proposed to interpret this cavity-free lasing \cite{Mysyrowicz2019,Xu2015,Richter2020}. Noting that the pump pulse at 800~nm is quasi-resonant with intermediate ionic level A (A$^2\Pi_u$), a first model assumes that fast depletion of the ionic ground state X leads to a population inversion between levels B and X \cite{Xu2015,Li2019,Ando2019}. However, this model does not explain the retarded emission of the lasing signal nor its dependence with gas pressure. A second model assumes no electronic population inversion and attributes the lasing to an inversion of rotational states of levels B and X \cite{Richter2020}. However, it ignores the role of pressure, of resonant intermediate level A and the effects of propagation of the lasing signal through the sample. We have recently developed another model that does not require electronic or rotational population inversion between initial state B and final state X for the initiation of lasing \cite{Mysyrowicz2019,Tikhonchuk2020}. This third interpretation relies on the observation of long-lived coherent polarizations X-A and X-B subsisting in the system well after the passage of the pump pulse \cite{Mysyrowicz2019}. Coupling between these polarizations in a V-scheme arrangement (see Figure 1 in Ref.~\cite{Mysyrowicz2019}) leads to the possibility of amplification of the B-X transition in the absence of population inversion ($n_B < n_X$). In this case, a softer condition for amplification requires $n_B > n_A$. Simulations based on Maxwell-Bloch equations, discussed in Ref.~\cite{Mysyrowicz2019} and in more details in Ref.~\cite{Tikhonchuk2020} in which rotational effects are included, restitute well the measured temporal shape of the lasing emission and the lasing response as a function of gas pressure.

The gain can be expressed as \cite{Tikhonchuk2020}:

\begin{equation}
E_bx = E_bx^0\exp\left({\frac{\mu_{ax}\mu_{bx}E_{p}^{0}}{3\hbar\mid{\omega_{ax}-\omega_{p}}\mid{}}\sqrt{\frac{N_i(n_B-n_A)\omega_{bx}\tau{z}}{c\epsilon_0\hbar}}}\right)   
\label{equation1}
\end{equation}

where E$_{bx}^0$ and E$_{bx}$ are the fields for the B-X transition at the beginning and at the end of the plasma column, $\omega_{ax}$ and $\omega_{bx}$ are the transition frequencies, $\mu_{ax}$=0.25~at.u. and $\mu_{bx}$=0.75~at.u. are the corresponding dipole moments, $\omega_p$ and $E_p^0$ are the frequency and amplitude of the laser post-pulse, z is the length of the plasma column, $\tau=t-z/c$ is the co-propagating time, $N_i$ is the density of N$_2^+$ ions, $\epsilon_0$ is the vacuum permittivity and $\hbar$ is the Planck constant. According to our model that includes the five resonant electronic levels B ($\nu=0$), X ($\nu=0$), X ($\nu=1$), A ($\nu=2$) and A ($\nu=3$), the repartition of ionic populations at the end of a 40~fs pump pulse with $2.5\times10^{14}~W/cm^2$ peak intensity is the following: $n_X (\nu=0)=31\%,~n_X (\nu=1) = 21\%,~n_B=19\%,~n_A (\nu=2)=16\%$~and~$n_A (\nu=3)=13\%$. At the end of the lasing signal at 391.4~nm, the population at level B has decreased by 2\% and the population at level A has increased by 30\% so that the lasing emission has been interrupted because condition $n_A<n_B$ was no more fulfilled.

\begin{figure}[!ht]
\centering
\includegraphics[width=10cm]{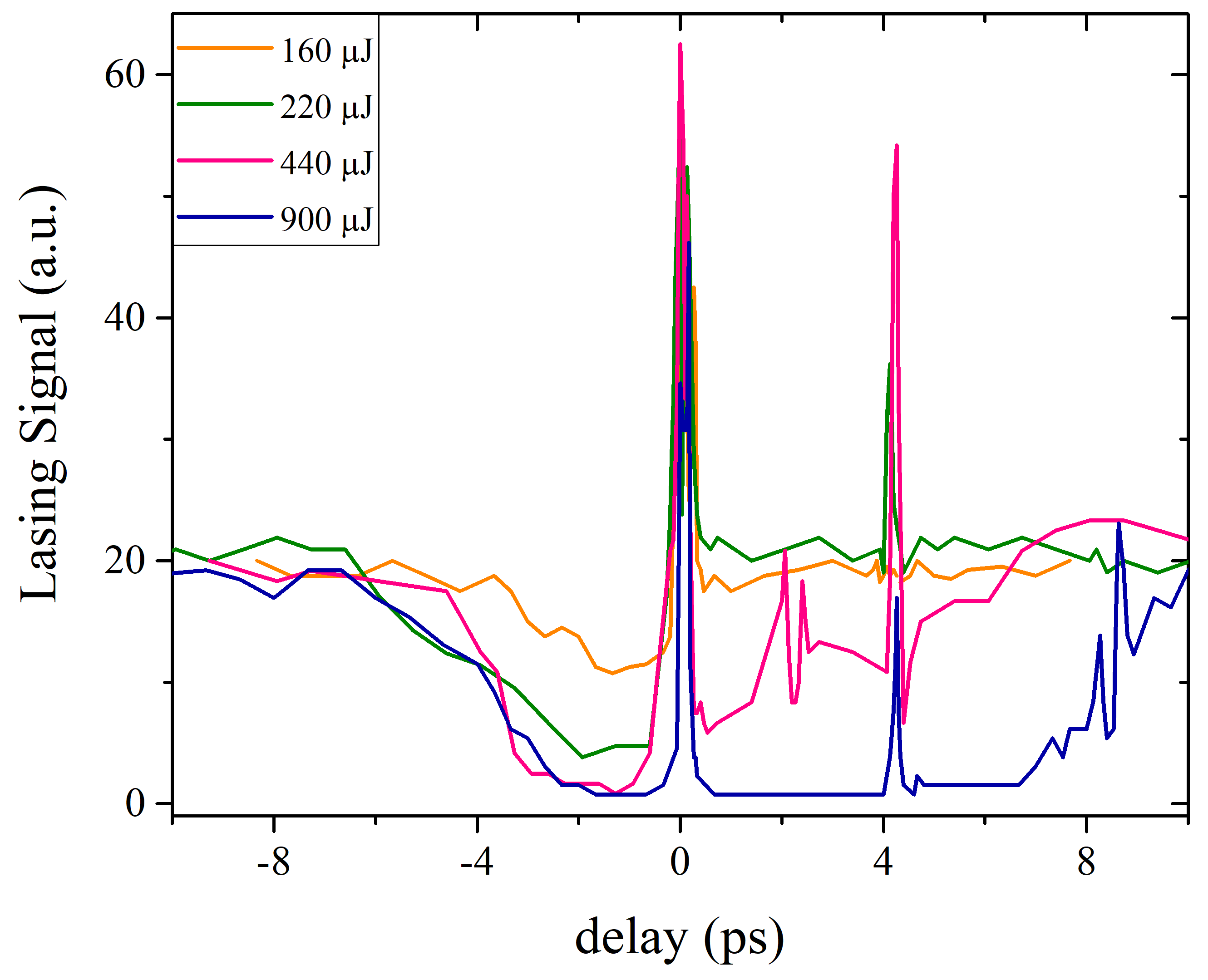}
\caption{Time-integrated lasing signal obtained with consecutive pump and control pulses. Pump pulse has 1.1 mJ energy, control pulse energy is indicated in the legend. N$_2$ gas pressure is ~100 mbar.}
\label{figure2}
\end{figure}

A clue to the origin of the quenching effect due to a second pump pulse is given in Figure \ref{figure2}. The lasing signal is shown as a function of delay between 800~nm pump and control pulses, for different energies, and consequently intensities, of the weaker control pulse, starting with an intensity below that required for high field ionization of the nitrogen gas. Positive delay $d$t in Figure \ref{figure2} corresponds to the weaker control pulse arriving first in the gas chamber. Inspecting first the lowest control pulse intensity, a reduction of lasing signal (due solely to the more intense pump pulse) appears only for $dt<0$, when the weaker control pulse follows the stronger pump pulse, while the amount of lasing for $dt>0$ is not affected. It indicates that the lasing partial quenching corresponds to an interaction of the control pulse with the plasma generated by the first IR pulse. Noting that the pump pulse wavelength at 800~nm is quasi-resonant with the X($\nu=0)\rightarrow$~A($\nu=2$) transition, we interpret the lasing erasing effect as due to a fast transfer of population from X to A, interrupting the coherent emission when the condition $n_A=n_B$ is reached. Obviously, the quenching can only appear for the duration of the lasing signal. According to Ref. \cite{Liu2015} and to the theoretical model of Ref. \cite{Tikhonchuk2020} the duration of the B-X lasing is about 10~ps for a gas pressure of 30~mbar, and it decreases to 4~ps when the pressure increases to 100~mbar.
Figure \ref{figure2} also shows how the lasing signal depends on the intensity of the control 800~nm pulse. For $dt<0$, the duration of quenching does not change appreciably, because it is essentially determined by the lasing duration of the stronger pulse, as discussed above. By contrast, for $dt>0$ we observe a longer duration at the highest control pulse intensities. This can be understood by the fact that the transfer of population X$\rightarrow$A responsible for quenching can now occur during the plasma lifetime created by the control pulse, which is longer than the lasing duration. We have verified that the quenching duration increases also for $dt<0$ when the collected lasing signal has a significant contribution from both pulses.
\par
\begin{figure}[!ht]
\centering
\includegraphics[width=10cm]{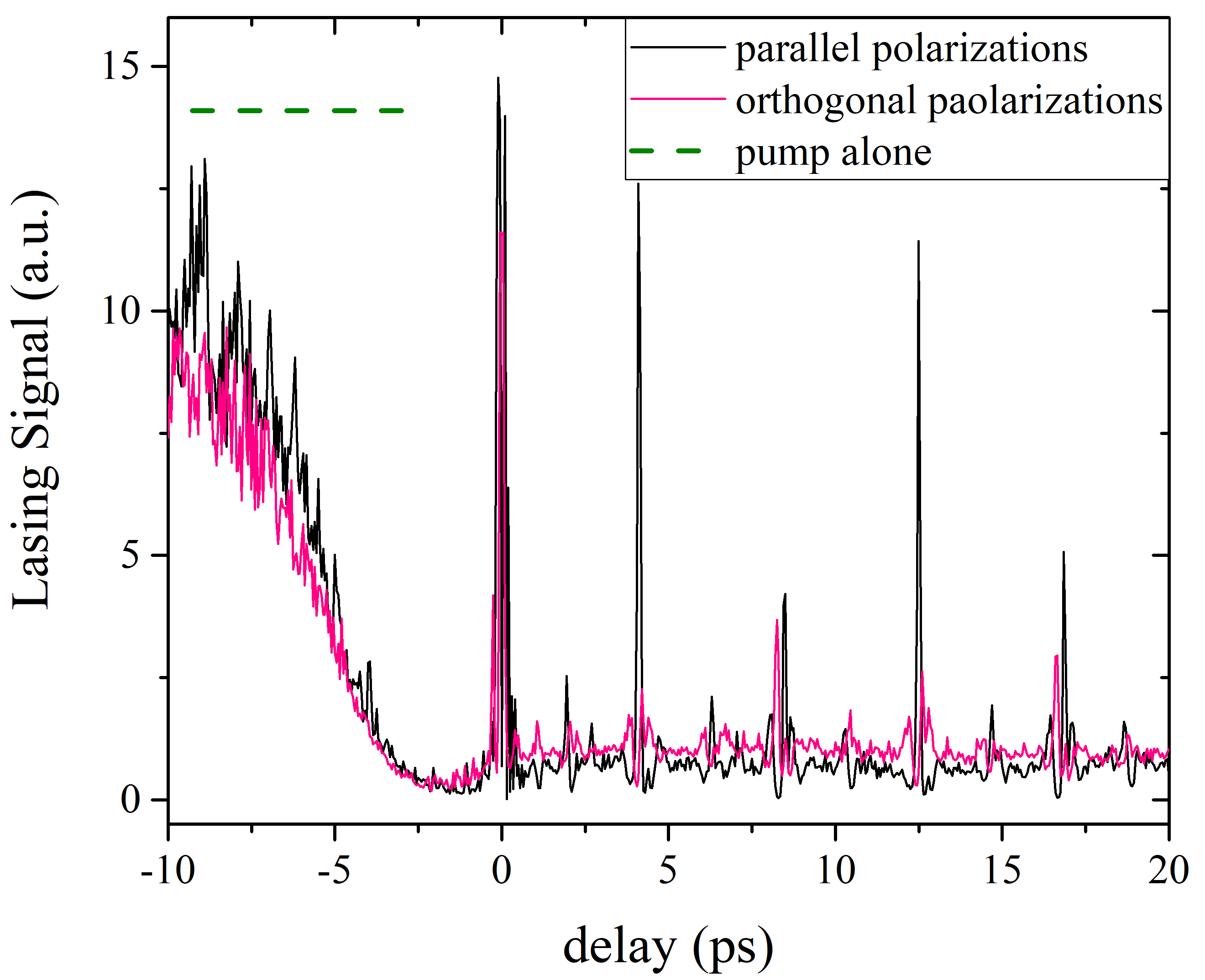}
\caption{Pink curve: Time-integrated lasing signal as a function of delay between consecutive pump and control pulses with parallel (in black) and orthogonal (in pinc) polarizations. Pulse energies are 1.2~mJ and 0.55 mJ. The dashed horizontal cyan line represents the magnitude of the lasing signal with 1.2 mJ pump pulse only. Maxima (minima) in the pinc and black curves correspond to an alignment of X ions parallel (perpendicular) to the first laser pulse field (see main text).}
\label{figure3}
\end{figure}

In addition to the quenching effect, one notices in Figure \ref{figure2} the appearance of sharp emission maxima and minima for $dt>0$. This effect is shown in more details in Figure \ref{figure3} for pump and control pulses having parallel and orthogonal polarizations. To explain the appearance of lasing signals during these short time intervals, two additional, well documented effects must be taken into account: the preparation of coherent rotational wave packets by the first pulse and the X$\rightarrow$A selection rule that forbids (allows) the X$\rightarrow$A transition when the field is parallel (perpendicular) to the molecular ion axis. A coherent rotational wave packet is induced in neutral nitrogen molecules by an intense short laser pulse that aligns them partially along the laser field. Because of inertia, a first alignment occurs $\sim$100~fs after the laser peak. Spontaneous revivals of rotational wave packets and concomitant realignment of neutral molecules occur with a specific period $T=1/4Bc=4.166$~ps, where $B=2$~cm$^{-1}$ is the rotational constant of N$_2$ and $c$ the speed of light, first along the laser field of the first pulse and then alternatively parallel and perpendicular to it \cite{Richter2020,Ripoche1997,Zeng2009}.

Let us examine the case of a second pulse with a polarization parallel to the first pulse impinging with a delay corresponding to a revival. Because of the anisotropic polarizability of N$_2$, during the short time $\sim100$~fs of the revival, the second pulse preferentially ionizes neutral molecules that are aligned along the first laser field. Therefore all ions, which are formed around the peak of the pulse predominantly in state X, become themselves aligned along the first pulse field. The quenching effect is inoperative for this alignment of the ions since the X$\rightarrow$A transition is forbidden when the electric field is parallel to the molecule axis. This synchronization explains the appearance of a lasing signal for this particular time delay. This is an efficient, phase-matched process, in the sense that the second pulse always encounters this favorable alignment of ionized molecules during its propagation through the sample. The recurrence time of lasing revivals shown in Figure \ref{figure3} is in excellent agreement with the time of partial alignment of the neutral molecules along the laser field calculated in reference \cite{Richter2020}. It is worth stressing that ions are responsible for the appearance of the lasing peaks even though the recurrence period corresponds to rotation of neutral molecules.
When the ions are aligned perpendicular to the first pulse field, a sign reversal is observed, lasing minima replacing lasing maxima. The same explanation holds. For molecules aligned perpendicular to the field of the first pulse, the X$\rightarrow$A transition is now allowed (the second pulse field is perpendicular to the molecules axis). The quenching effect is therefore reinforced. The positions of quenching minima are again in excellent agreement with the calculated alignments of molecules perpendicular to the first field. Finally, we have verified experimentally that minima replace maxima and vice-versa when pump and control pulses have orthogonal polarizations, a result explained by the same reasoning (see Figure \ref{figure3}).
\par
\begin{figure}[!ht]
\centering
\includegraphics[width=10cm]{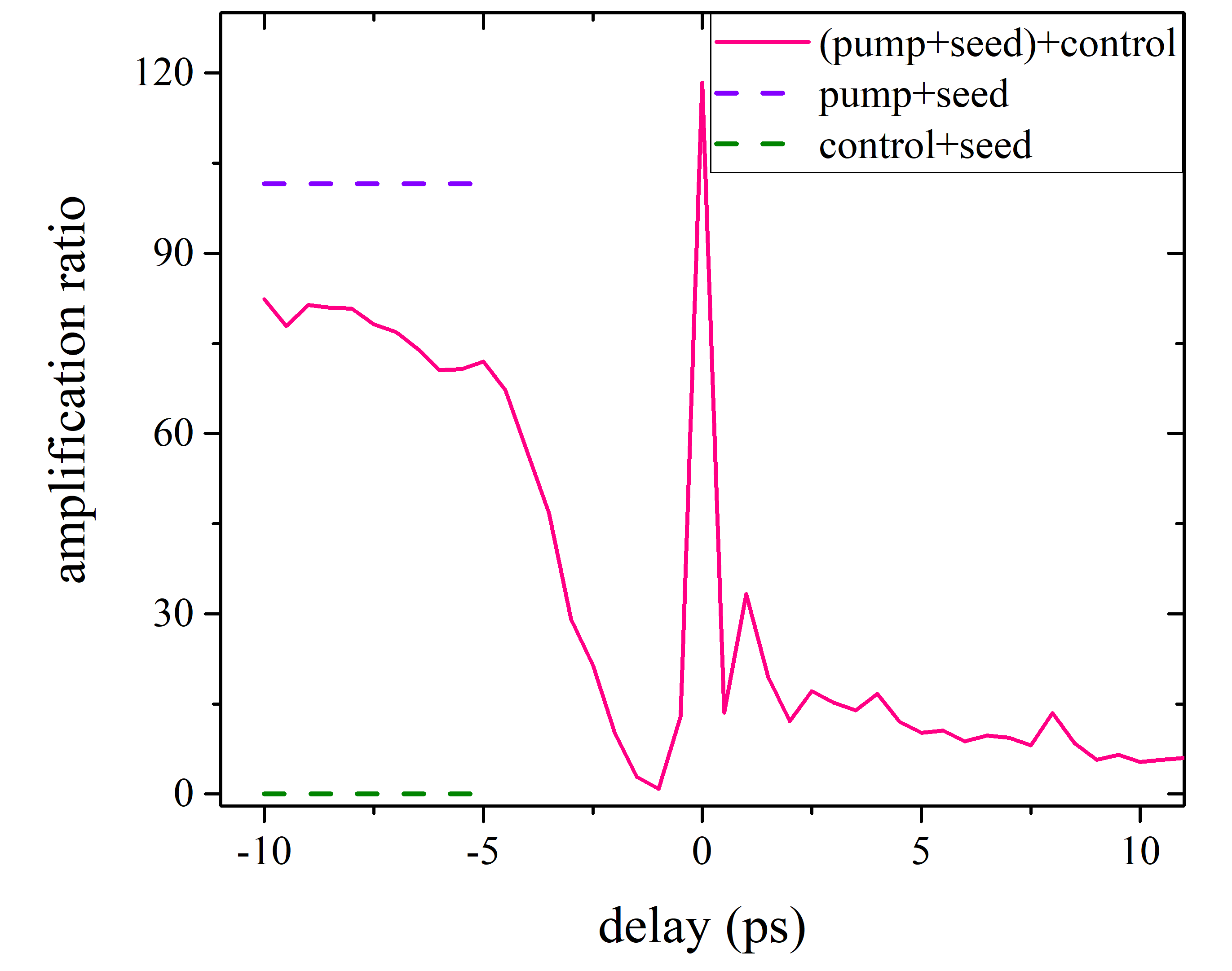}
\caption{Pink curve: Time-integrated amplification ration as a function of delay between a fixed pump and a scanned control pulse at 800~nm. Pump pulse energy is 0.8~mJ. Control pulse energy is 0.7~mJ. Pulse duration is 45~fs. An external seed pulse around 391~nm with 200 fs duration is injected 300~fs after the pump pulse. Amplification ratio refers to the ratio between amplified lasing signal and injected external seed, self-seeded lasing intensity being negligible. Gas pressure is 30 mbar. Violet dash horizontal line represents the magnitude of the laser amplification for the 0.8 mJ pump pulse when the weak seed at 391 nm is injected. Green dash horizontal line gives the amplification from the control pulse (see main text for more details).}
\label{figure4}
\end{figure}

So far we have only discussed the erasing effect of self-seeded lasing in N$_2^+$. As mentioned before, an external femtosecond seed pulse around 391~nm amplifies the lasing signal by several orders of amplitude. It is interesting to verify if the quenching is still effective in this case. To answer this question experimentally, an external seed around 391~nm with $\sim200$~fs duration was generated in a mismatched BBO-crystal and injected $\sim300$~fs after the 800~nm pump pulse, leading to amplification of the lasing signal by a factor $\sim100$. By comparison, the lasing signal from the control pulse remained negligibly small. As shown in Figure \ref{figure4}, a significant erasing of lasing occurred even in this case for both $dt>0$ and $dt<0$. In particular, lasing totally vanished for $dt=-1$~ps. The poorer scan resolution (500~fs instead of 50~fs scan steps in Figure \ref{figure3}) prevents the clear observation of lasing revivals. To sum up, our arguments about the origin of the lasing suppression remain valid in the presence of an external seed.

\section{Conclusion}
In conclusion, we have reported the nearly total quenching of cavity-free lasing of N$_2^+$ by two consecutive femtosecond laser pulses at 800~nm wavelength. This effect has its root in the interference between amplitudes of transitions between levels X$^2\Sigma_g^+$, A$^2\Pi_u$ and B$^2\Sigma_u^+$ of N$_2^+$ in a V-arrangement. The V-scheme of lasing without population inversion explains naturally the origin of this double pulse erasing process. Quenching is due to a transition from ground X to first excited state A of singly ionized nitrogen molecules induced by the second pump pulse, leading to an abrupt stop of the lasing when the populations of levels A and B become equal. The same process also explains the subtle behavior of lasing for  delays between the two pump pulses corresponding to coherent wave packet revivals. We finally note that the presented results seem in contradiction with other proposed models \cite{Xu2015,Li2019,Ando2019,Richter2020}. In the model based on population inversion between B and X, which has the merit to have pointed out first the role of intermediate level A, a second pulse at 800~nm with a short delay should further deplete level X in a X-A transition and consequently should lead to an increase, not a decrease of lasing signal. The rotation inversion model predicts gain windows that are slightly shifted in time from the lasing revival times observed here. It does not explain the quenching effect. In our opinion rotation level inversion is not at the origin of lasing in N$_2$ although the model is certainly valuable in explaining a reinforcement of the lasing at certain delays.

\bigskip
\bigskip
\noindent\textbf{Acknowledgment.}~The work has been partially supported by CNRS, by Project LQ1606 from the Czech National Programme of Sustainability II  and by the National Natural Science Foundation of China (Grants No. 11574213), Innovation Program of Shanghai Municipal Education Commission (Grant No. 2017-01-07-00-07-E00007), Shanghai Municipal Science and Technology Commission (No. 17060502500).

{}
\newpage

\section*{figures}

\end{document}